\documentclass[11pt]{article}

\usepackage{graphicx}
\usepackage{amssymb}
\usepackage{float}
\usepackage{chet}

\newcommand{\CO}{\mathcal{O}}
\newcommand{\CN}{\mathcal{N}}
\newcommand{\CF}{\mathcal{F}}
\newcommand{\CH}{\mathcal{H}}
\newcommand{\lsp}{\hspace{1pt}}
\newcommand{\<}{\langle}
\renewcommand{\>}{\rangle}

\date{September 2015}

\title{Exploring the Minimal 4D $\titlemath{\CN=1}$ SCFT}

\author{David Poland$^{a,\lsp b}$ and Andreas Stergiou$^{a}$
\emails{(\href{mailto:david.poland@yale.edu}{david.poland},
\href{mailto:andreas.stergiou@yale.edu}{andreas.stergiou})@yale.edu}}

\affiliation{$^{a}$Department of Physics, Yale University, New Haven, CT
06520, USA\\ $^{b}$School of Natural Sciences, Institute for Advanced
Study, Princeton, NJ 08540, USA}

\abstract{We study the conformal bootstrap constraints for 4D $\CN=1$
superconformal field theories containing a chiral operator $\phi$ and the
chiral ring relation $\phi^2=0$. Hints for a minimal interacting SCFT in
this class have appeared in previous numerical bootstrap studies. We
perform a detailed study of the properties of this conjectured theory,
establishing that the corresponding solution to the bootstrap constraints
contains a $\text{U}(1)_R$ current multiplet and estimating the central
charge and low-lying operator spectrum of this theory.}

\begin{document}

\maketitle

\newsec{Introduction}
The conformal bootstrap has emerged as a powerful tool for studying
conformal field theories (CFTs) in $D>2$, with numerous applications. In
recent years it has allowed us to learn precise quantitative information
about known strongly-interacting CFTs, such as the 3D
Ising~\cite{ElShowk:2012ht, El-Showk:2014dwa, El-Showk:2013nia,
Kos:2014bka,Simmons-Duffin:2015qma}, $\text{O}(N)$
vector~\cite{Kos:2013tga, Nakayama:2014yia, Bae:2014hia, Chester:2014gqa,
Nakayama:2014lva,Nakayama:2014sba, Shimada:2015gda, Kos:2015mba}, and
Gross--Neveu models~\cite{Iliesiu:2015qra}, known 3D
$\CN=2$~\cite{Bobev:2015vsa, Bobev:2015jxa, Chester:2015qca} and
$\CN=8$~\cite{Chester:2014fya, Chester:2014mea} SCFTs, 4D
$\CN=2$~\cite{Beem:2013sza, Beem:2014rza, Beem:2014zpa} and $\CN=4$
SCFTs~\cite{Beem:2013qxa, Alday:2013opa, Alday:2014qfa, Alday:2014tsa}, and
the mysterious 6D $(2,0)$ SCFTs~\cite{Beem:2014kka, Beem:2015aoa}.
Moreoever, because it probes the full space of CFTs without reference to
any particular microscopic description, the conformal bootstrap is also a
powerful tool for discovering new, previously unknown theories.

Hints for a possible new 4D CFT with $\CN=1$ supersymmetry appeared
in~\cite{Poland:2011ey} (building on earlier
studies~\cite{Rattazzi:2008pe,Rychkov:2009ij, Caracciolo:2009bx,
Poland:2010wg, Rattazzi:2010gj, Rattazzi:2010yc, Vichi:2011ux}),
manifesting as a kink in general bounds on the scaling dimension of the
leading non-chiral scalar in the OPE $\bar{\phi} \times \phi$, where $\phi$
is a chiral operator. This coincided with the disappearance of a lower
bound on the chiral operator OPE coefficient $\phi \times \phi \sim
\lambda_{{\phi}^2} \phi^2$, allowing this coefficient to vanish precisely
at this dimension. Moreover, it was established in~\cite{Bobev:2015jxa}
that a similar feature appears at all $2 \leq D \leq 4$ in SCFTs with four
supercharges, where as $D \rightarrow 2$ it merges with the 2D $\CN=2$
minimal model. The absence of the $\phi^2$ operator in $D=3,4$ could also
be seen more directly in the approximate solutions to crossing symmetry
reconstructed in~\cite{Bobev:2015jxa}.

However, the correct interpretation of these features in both $D=3,4$ is
not yet understood. Based on their similarity to features that are known to
coincide with the 3D Ising~\cite{ElShowk:2012ht, El-Showk:2014dwa,
El-Showk:2013nia, Kos:2014bka,Simmons-Duffin:2015qma} and 3D $\text{O}(N)$
vector models~\cite{Kos:2013tga, Kos:2015mba}, it is tempting to conjecture
the existence of a family of new SCFTs. In this work we study the 4D
$\CN=1$ version of these kinks in greater detail, exploring the properties
of the theory that we conjecture to live there.

We will establish several properties of this conjectured theory using the
conformal bootstrap conditions for the correlator $\<\bar{\phi} \phi
\bar{\phi} \phi\>$, building on the earlier results
of~\cite{Poland:2011ey,Bobev:2015jxa}. First we establish directly that
assuming the chiral ring condition $\phi^2 = 0$ imposes a sharp lower bound
$\Delta_{\phi} \geq 1.415$. In particular we exclude the possibility that
$\Delta_{\phi} = \sqrt{2}$. Second, after imposing the chiral ring
condition we place a bound on the leading spin-1 superconformal primary and
find that it forces the existence of a $\text{U}(1)_R$ current multiplet
when the lower bound on $\Delta_{\phi}$ is saturated.

Having established that this putative theory contains a $\text{U}(1)_R$
current multiplet (whose descendant is the stress-energy tensor), we
proceed to compute general lower and upper bounds on the conformal central
charge for SCFTs with $\phi^2 = 0$. The upper bounds are somewhat dependent
on the gap until the next spin-1 primary, but for all gaps the lower and
upper bounds merge at the minimal value of $\Delta_{\phi}$. We estimate
that this minimal theory has $c/c_{\text{free}} \simeq 8/3$ where
$c_{\text{free}}$ is the central charge of a free chiral multiplet. We also
make preliminary determinations of the OPE coefficient of the
$\bar{\phi}\phi$ operator, the dimensions of the second scalar and spin-1
superconformal primaries, and the dimension of the leading spin-2
superconformal primary.

In the present work we have not yet found a set of gap assumptions that
isolate this solution, i.e.\ we do not yet see islands analogous to what
was found in~\cite{Kos:2014bka,Simmons-Duffin:2015qma,Kos:2015mba}. For
this we anticipate that we will need to consider a larger system of
correlators containing both the $\phi$ and $\bar{\phi}\phi$ operators.
However, in our current setup we can already uncover a lot of information
about this theory and we hope that the results of this paper are a useful
step towards identifying the nature of this mysterious 4D $\CN=1$ SCFT.

\newsec{Results}
In this work we study the correlator $\<\bar{\phi} \phi \bar{\phi} \phi\>$
where $\phi$ is a chiral operator in a 4D $\CN=1$ SCFT, similar to what was
done in~\cite{Poland:2011ey}. Crossing symmetry of this correlator leads to
the sum rules
\eqn{\sum_{\CO \in \bar{\phi} \times \phi} |\lambda_{\CO}|^2
\begin{pmatrix}
  \CF_{\Delta,\ell}(z,\bar{z}) \\
  \tilde{\CF}_{\Delta,\ell}(z,\bar{z}) \\
  \tilde{\CH}_{\Delta,\ell}(z,\bar{z})
\end{pmatrix}
+ \sum_{\CO \in \phi \times \phi} |\lambda_{\CO}|^2
\begin{pmatrix}
  0 \\
  F_{\Delta,\ell}(z,\bar{z}) \\
  -H_{\Delta,\ell}(z,\bar{z})
\end{pmatrix} = 0\,,}[]
where the functions of the conformal cross-ratios $z$ and $\bar{z}$ that
appear are related to conformal and superconformal blocks and defined
in~\cite{Poland:2011ey}.\footnote{Additional results and formalism for
setting up the 4D $\CN=1$ superconformal bootstrap has e.g.\ been developed
in~\cite{Park:1997bq, Osborn:1998qu, Poland:2010wg, Vichi:2011ux,
Fortin:2011nq,Goldberger:2011yp, Goldberger:2012xb, Khandker:2012pa,
Berkooz:2014yda, Khandker:2014mpa, Fitzpatrick:2014oza, Kumar:2014uxa,
Li:2014gpa, Bobev:2015jxa}.} In general we assume that the superconformal
primary operators $\CO$ in the first sum satisfy the unitarity bound
$\Delta_{\CO} \geq \ell+2$~\cite{Flato:1983te, Dobrev:1985qv}, while the
even-spin operators in the second sum may either be conformal primaries in
BPS multiplets with $\Delta_{\CO} = 2\lsp\Delta_{\phi} + \ell$, or
conformal primaries in unprotected multiplets satisfying the unitarity
bound $\Delta_{\CO} \geq |2\lsp\Delta_{\phi}-3| + 3 +
\ell$~\cite{Poland:2010wg, Vichi:2011ux}.

As described in~\cite{Poland:2011ey}, in order to rule out assumptions
about the spectrum we can look for a 3-vector of functionals $\vec{\alpha}$
that when applied this sum rule leads to a contradiction. In
particular, if the functional is $>0$ on the identity operator contribution
and $\geq 0$ on all other possible contributions, then the sum rule can
never be satisfied. Alternatively, by normalizing the functional on the
contribution of a particular operator $\CO_0$ and extremizing the action on
the identity operator, we can obtain upper or lower bounds on the OPE
coefficient of $\CO_0$. We apply this logic below to obtain bounds on
operator dimensions and OPE coefficients, using
SDPB~\cite{Simmons-Duffin:2015qma} to solve the relevant optimization
problem after phrasing it in the language of semi-definite programming. The
functional search space is governed by the parameter $\Lambda$, where each
component  $\alpha_i$ is a linear combination of $\frac12\left\lfloor
\frac{\Lambda+2}{2}\right\rfloor \left(\left\lfloor \frac{\Lambda+2}{2}
\right\rfloor + 1\right) $ independent nonvanishing derivatives $\alpha_i
\propto \sum_{m,n} a_{mn} \partial_z^m \partial_{\bar{z}}^n
\big|_{1/2,1/2}$ with $m+n \leq \Lambda$.

\begin{figure}[!t]
  \begin{center}
    \includegraphics{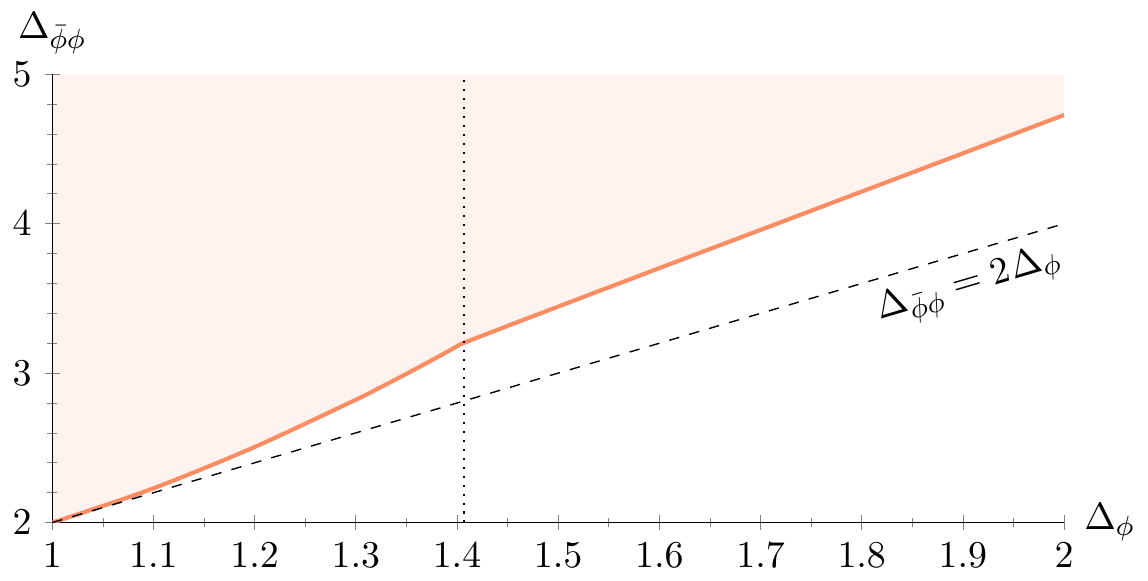}
  \end{center}
  \vspace{-11pt}
  \caption{Upper bound on the allowed dimension of the operator
  $\bar{\phi}\phi$ (the leading relevant nonchiral scalar singlet) as a
  function of the dimension of $\phi$. The generalized free theory dashed
  line $\Delta_{\bar{\phi}\phi}=2\hspace{0.5pt}\Delta_{\phi}$ is also
  shown. The shaded area is excluded.  If we assume that $\phi^2$ is not in
  the spectrum then everything to the left of the dotted line at
  $\Delta_\phi=1.407$, which is the position of the kink, is excluded. Here
  we use $\Lambda=21$.}
  \label{fig:dim_phibphi}
\end{figure}

First, we reproduce the general upper bound on the dimension of the leading
unprotected scalar operator $\Delta_{\bar{\phi}\phi}$, finding precise
agreement with~\cite{Poland:2011ey}. This bound is shown in
Fig.~\ref{fig:dim_phibphi}.
At $\Lambda=21$ there is a mild kink in this upper bound around
$\Delta_{\phi} \simeq 1.407$. We show how this position changes as we
increase the search space of the functional in a later plot. Note that any
theory saturating this bound necessarily does not contain any scalar
superconformal primaries of dimension 2, i.e. $\phi$ cannot be charged
under any global symmetries.

Next we recompute this bound imposing the additional condition that the
chiral $\phi^2$ operator does not appear in the $\phi \times \phi$ OPE.
This condition has the effect of excluding all points to the left of the
dotted vertical line in Fig.~\ref{fig:dim_phibphi}. The region to the right
remains the same.  In other words, it imposes the strict lower bound
$\Delta_{\phi} \geq 1.407$, causing the mild kink to turn into a sharp
corner.

\begin{figure}[!t]
  \begin{center}
    \includegraphics{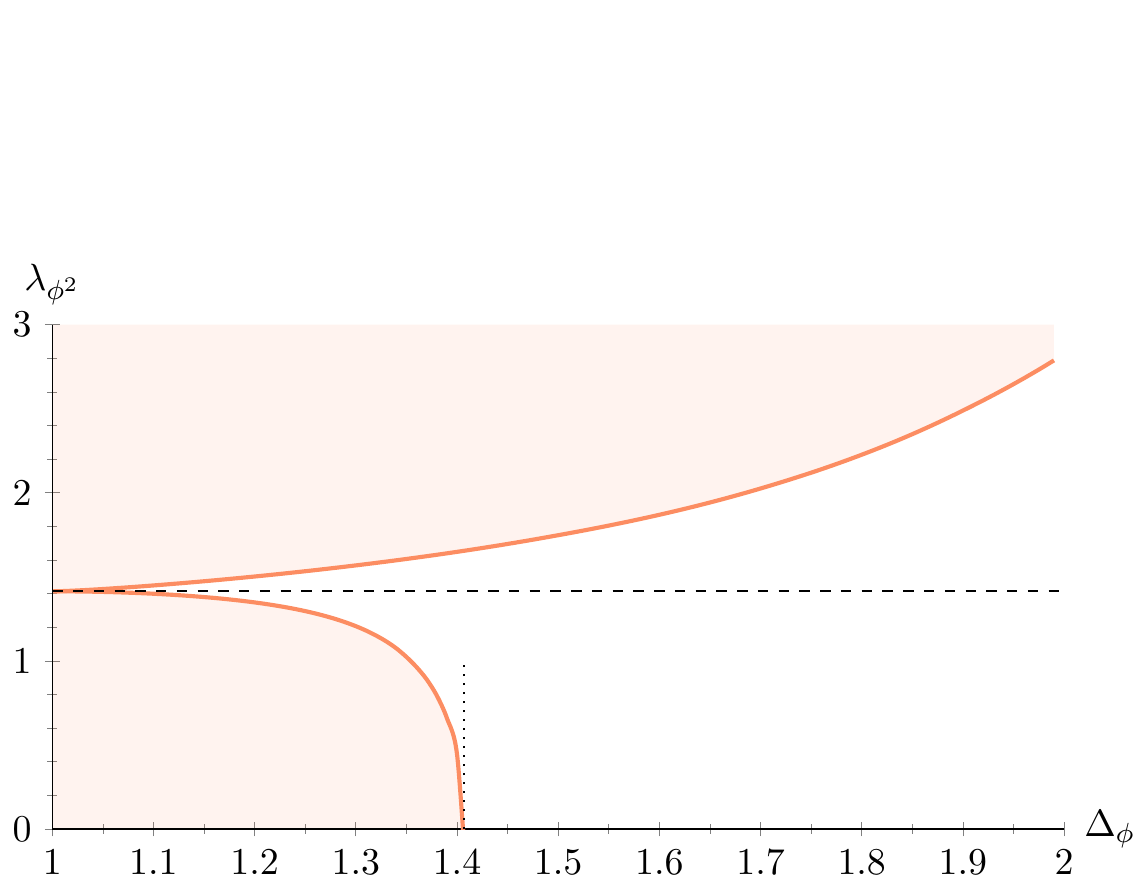}
  \end{center}
  \vspace{-12pt}
  \caption{Lower and upper bounds on the OPE coefficient of the operator
  $\phi^2$ in the $\phi\times\phi$ OPE.  The vertical dotted line is at
  $\Delta_\phi=1.407$ and the horizontal dashed line is at the free theory
  value $\lambda_{\phi^2}=\sqrt{2}$. The shaded area is excluded.  Here we
  use $\Lambda=21$.}
  \label{fig:phisq_upper_lower_bound}
\end{figure}

One can see that this had to be the case by considering bounds on the OPE
coefficient of the operator $\phi^2$, shown in
Fig.~\ref{fig:phisq_upper_lower_bound}.
The lower bound on $\lambda_{\phi^2}$ disappears exactly at
$\Delta_\phi=1.407$. Thus, Fig.~\ref{fig:phisq_upper_lower_bound} makes it
clear that if we demand $\lambda_{\phi^2}=0$, implying that $\phi^2$ is not
in the spectrum, then all points to the left of $\Delta_\phi=1.407$ must be
excluded. Our general bound is also compatible with the results
of~\cite{Bobev:2015jxa}, which found that the $\phi^2$ operator was absent
in approximate solutions to crossing symmetry living on the boundary of the
allowed region to the right of the kink.

\begin{figure}[!t]
  \begin{center}
    \includegraphics{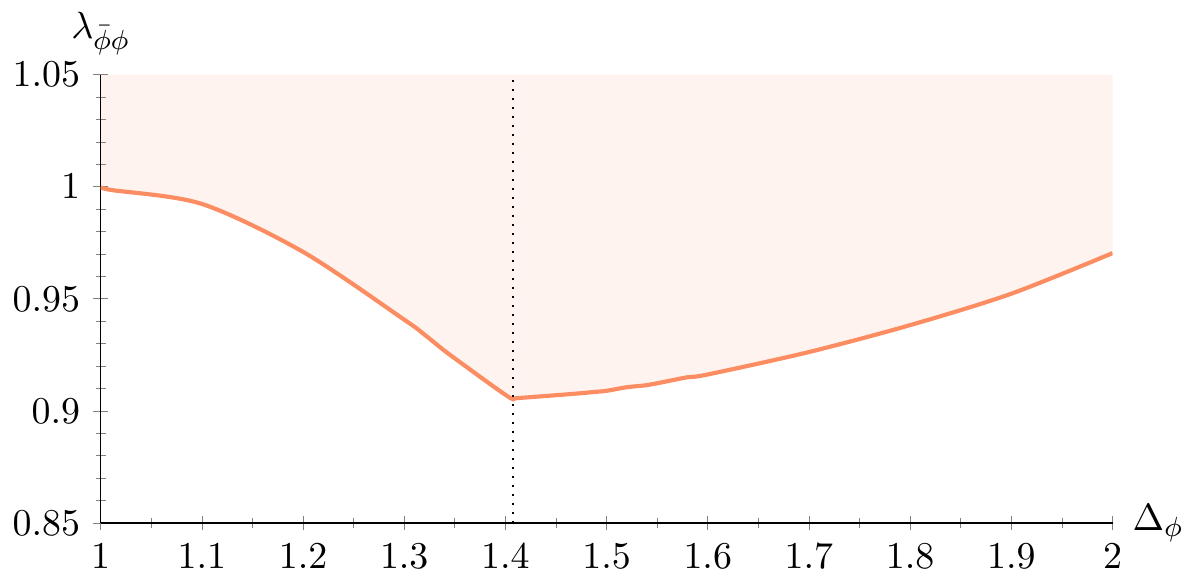}
  \end{center}
  \vspace{-12pt}
  \caption{Upper bound on the OPE coefficient of an operator
  $\bar{\phi}\phi$ with dimension
  $\Delta_{\bar{\phi}\phi}^{(\text{bound})}$ as a function of the dimension
  of $\phi$. Here we do not assume that $\bar{\phi}\phi$ is the scalar with
  the lowest dimension in the OPE $\bar{\phi}\times\phi$. The shaded area
  is excluded. In this plot we use $\Lambda=21$.}
  \label{fig:phibphi_ope_coeff}
\end{figure}

In Fig.~\ref{fig:phibphi_ope_coeff} we show an upper bound on the OPE
coefficient $\lambda_{\bar{\phi}\phi}$ of an operator whose dimension
saturates the bound in Fig.~\ref{fig:dim_phibphi}.
Without any additional assumptions the upper bound attains a minimum at
precisely the location of the kink, occuring at $\lambda_{\bar{\phi}\phi}
\simeq 0.905$. If we further impose the absence of $\phi^2$, then all
points to the left of the dotted vertical line in
Fig.~\ref{fig:phibphi_ope_coeff} are excluded.

Next we would like to ask the question: if there is an SCFT living near the
kink with the chiral ring relation $\phi^2=0$, does it contain a
stress-energy tensor? In other words, could it correspond to a local SCFT?
\begin{figure}[!t]
  \begin{center}
    \includegraphics{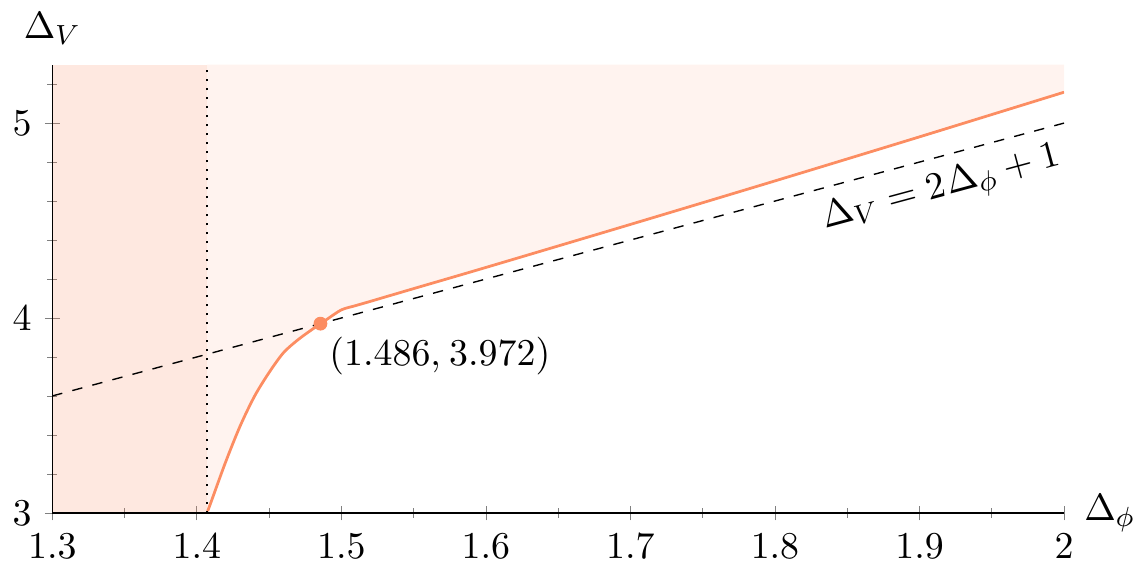}
  \end{center}
  \vspace{-12pt}
  \caption{Upper bound on the dimension of the leading superconformal
  primary vector operator in the OPE $\bar{\phi}\times\phi$ as a function
  of the dimension of $\phi$. The shaded area is excluded. Everything to
  the left of the vertical dotted line at $\Delta_{\phi}=1.407$ is excluded
  due to the assumption that there is no $\phi^2$ operator. The generalized
  free theory dashed line $\Delta_V=2\hspace{0.5pt}\Delta_{\phi}+1$ as well
  as its intersection with the bound are also shown.  In this plot we use
  $\Lambda=21$.}
  \label{fig:dim_vector_plot_with_assumption}
\end{figure}
In Fig.~\ref{fig:dim_vector_plot_with_assumption} we assume $\phi^2=0$ and
place an upper bound on the leading spin-1 superconformal primary $V$ in
the $\bar{\phi} \times \phi$ OPE, again at $\Lambda = 21$. We see that the
bound on $\Delta_V$ approaches 3 as $\Delta_{\phi}$ approaches its minimum
value. Thus, the $\text{U}(1)_R$ current multiplet $V_R$ is required to be
in the spectrum at this point.

Note that for sufficiently small $\Delta_{\phi}$ the bound excludes the
line that would correspond to a generalized free theory with $\Delta_V =
2\lsp\Delta_{\phi} +1$. This is natural, as our assumption that $\phi^2$ is
absent is not true in a generalized free theory. On the other hand, when
$\Delta_{\phi} \geq 3/2$, the contribution in the sum rule corresponding to
the chiral $\phi^2$ operator is identical to one contained in the
unprotected scalar contributions in $\phi \times \phi$. Thus, we expect the
generalized free line should be allowed for $\Delta_{\phi} \geq 3/2$. Here
we see that it crosses this line at $\Delta_{\phi} \sim 1.486$, compatible
with this expectation.

\begin{figure}[!t]
  \begin{center}
    \includegraphics{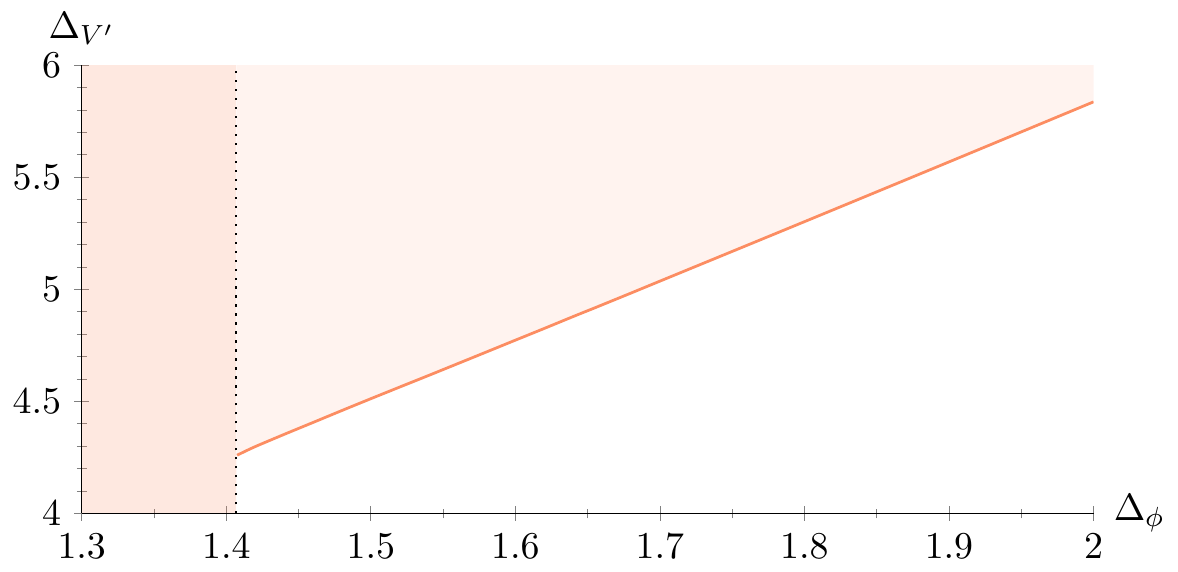}
  \end{center}
  \vspace{-12pt}
  \caption{Upper bound on the dimension of the second superconformal
  primary vector operator in the OPE $\bar{\phi}\times\phi$ as a function
  of the dimension of $\phi$, assuming that the first vector has dimension
  3. the shaded area is excluded.  everything to the left of the vertical
  dotted line at $\Delta_{\phi}=1.407$ is excluded due to the assumption
  that there is no $\phi^2$ operator.  In this plot we use $\Lambda=21$.}
  \label{fig:dim_second_vector_plot_with_assumption}
\end{figure}

Now that we have established the existence of a $\text{U}(1)_R$ current
multiplet, we can assume it to be in the spectrum and place an upper bound
on the second spin-1 operator $V'$. The result is shown in
Fig.~\ref{fig:dim_second_vector_plot_with_assumption}.  We see that
$\Delta_{V'} \lesssim 4.25$ at the minimum value of $\Delta_{\phi}$.

\begin{figure}[!t]
  \begin{center}
    \includegraphics{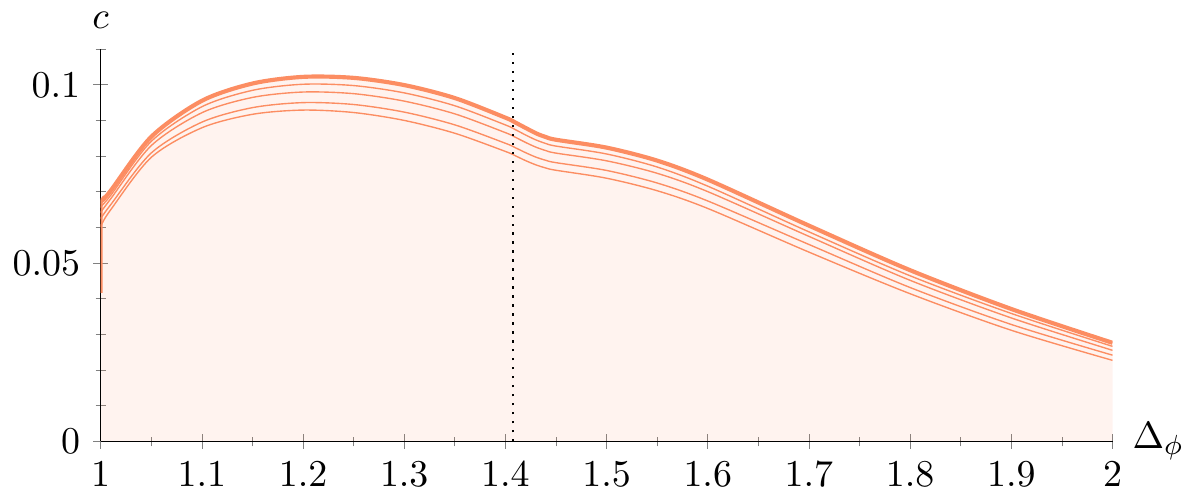}
  \end{center}
  \vspace{-11pt}
  \caption{Lower bound on the central charge as a function of the dimension
  of $\phi$. The shaded area is excluded.  For the strongest bound (thick
  line) we use $\Lambda=29$, while for the weaker bounds (thin
  lines) we use $\Lambda=21, 23, 25, 27$ (from bottom to top). The
  dotted line is at $\Delta_{\phi}=1.407$.}
  \label{fig:central_charge}
\end{figure}

We can also compute general lower bounds on the central charge $c$, using
that the OPE coefficient $\lambda_{V_R}^2 = \Delta_{\phi}^2/72\lsp c$. Here
our normalization is such that $c_{\text{free}} = 1/24$ for a free chiral
multiplet. Similar bounds were computed in~\cite{Poland:2011ey}.  Here
these bounds are shown in Fig.~\ref{fig:central_charge} for $\Lambda = 21,
23, \ldots, 29$. As in~\cite{Poland:2011ey}, these bounds drop very sharply
as $\Delta_{\phi} \rightarrow 1$ so as to be compatible with the free
theory value $c_{\text{free}} = 1/24$.

\begin{figure}[!t]
  \begin{center}
    \includegraphics{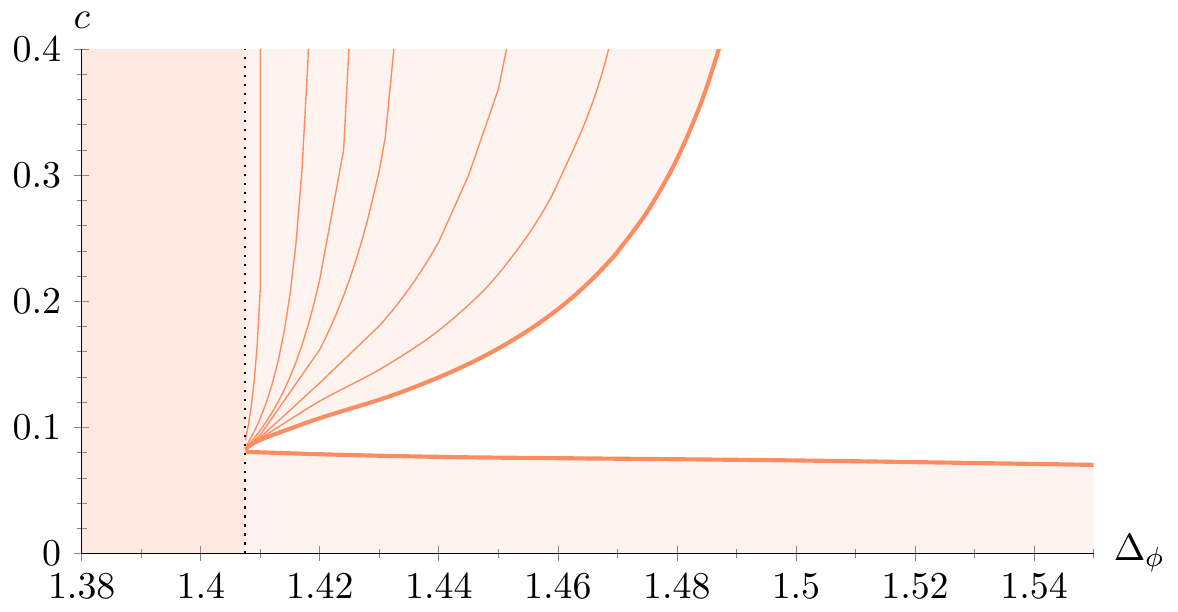}
  \end{center}
  \vspace{-12pt}
  \caption{Lower and upper bounds on the central charge as a function of
  the dimension of $\phi$, with the assumption that there is no $\phi^2$
  operator and all vector operators but the first one obey
  $\Delta_{V_{\text{other}}}\ge4.1$
  (thick upper bound line). The thinner upper bound lines correspond to
  $\Delta_{V_{\text{other}}}\ge3.1, 3.3, 3.5, 3.7, 3.9, 4$ (from left to
  right).  The shaded area is excluded.  In this plot we use $\Lambda=21$.}
  \label{fig:central_charge_upper_lower_bound}
\end{figure}

We can also impose a gap until the second spin-1 dimension $\Delta_{V'}$
and find upper bounds on $c$ for each value of the gap. These bounds are
shown in Fig.~\ref{fig:central_charge_upper_lower_bound} at $\Lambda = 21$,
where we have also imposed that there is no $\phi^2$ operator. We see that
the upper and lower bounds meet at the minimum value of $\Delta_{\phi}$,
essentially uniquely fixing the central charge at this point, with $c
\simeq .081$ at $\Lambda = 21$.

\begin{figure}[!t]
  \begin{center}
    \includegraphics{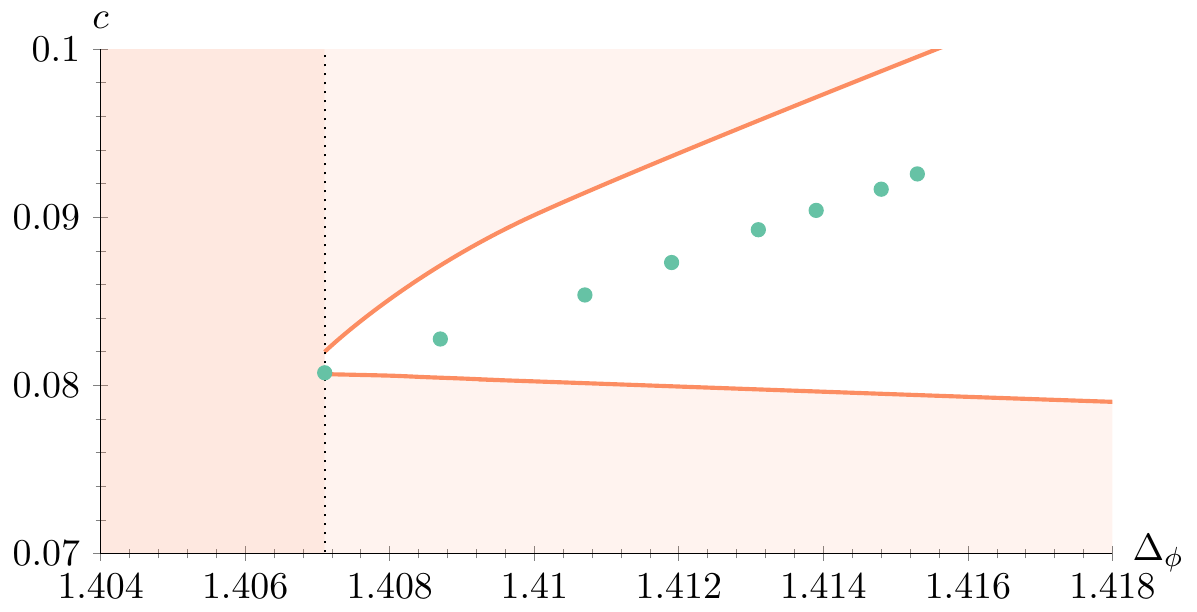}
  \end{center}
  \vspace{-12pt}
  \caption{Lower and upper bounds on the central charge as a function of
  the dimension of $\phi$, with the assumptions that there is no $\phi^2$
  operator and that all vector operators but the first one obey
  $\Delta_{V_{\text{other}}}\ge4.1$. The shaded area is excluded. Here we
  use $\Lambda=21$ for the bounds. The green points are allowed points
  closest to the corresponding lower bound for $\Lambda=21, 23, \ldots, 35$
  (from left to right).}
  \label{fig:central_charge_upper_lower_bound_zoomed}
\end{figure}

\begin{figure}[!t]
  \begin{center}
    \includegraphics[width=.48\textwidth]{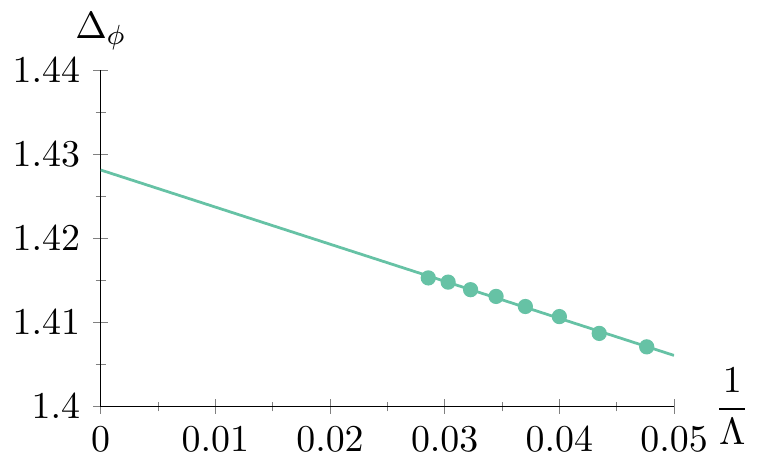}
    \hspace{.5cm}\includegraphics[width=.48\textwidth]{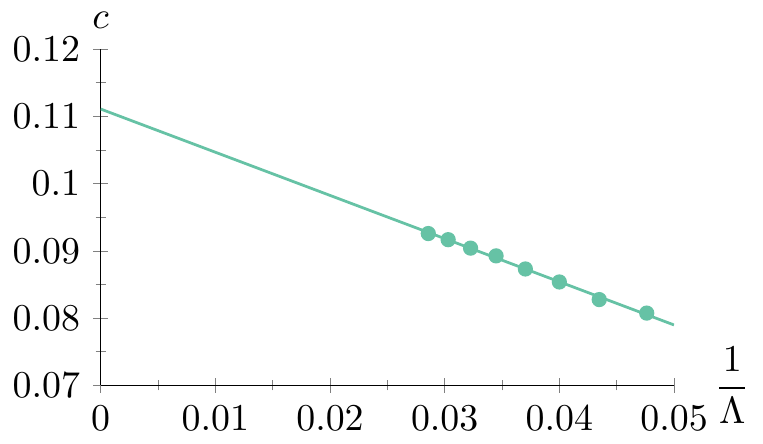}
  \end{center}
  \vspace{-12pt}
  \caption{Linear extrapolations of the position of the minimal value of
  $\Delta_{\phi}$ (assuming $\phi^2$ is absent) and the corresponding value
  of $c$ as a function of the inverse cutoff $1/\Lambda$.}
  \label{fig:extrapolations}
\end{figure}

On the other hand, as seen in Fig.~\ref{fig:central_charge}, our bounds
have not yet converged, so the location of this unique point in
$\{\Delta_{\phi}, c\}$ space will change somewhat at larger values of
$\Lambda$. We have explored the location of this point up to $\Lambda =
35$, shown in Fig.~\ref{fig:central_charge_upper_lower_bound_zoomed}.
Our strongest bound is $\Delta_{\phi} \geq 1.415$ at $\Lambda = 35$.
In this plot we also compare these points to the upper and lower bounds on
$c$ computed at $\Lambda = 21$ and $\Delta_{V'} \geq 4.1$ ($\Delta_{V'}
\geq 4.2$ seems to be excluded at $\Lambda = 35$). Unfortunately, the
location has not yet completely converged at $\Lambda = 35$, but there is a
striking linear relation between $\Delta_{\phi}$ and $c$, given
approximately by $c \approx 1.454\lsp \Delta_{\phi} - 1.965$.  Moreover, as
we increase $\Lambda$ the rate of convergence appears to be well-described
by a fit that is linear in $1/\Lambda$ (similar to the fit done
in~\cite{Beem:2015aoa}),
\eqn{\left\{\Delta_{\phi}(\Lambda), c(\Lambda) \right\} \approx
\left\{1.428 - \frac{0.441}{\Lambda}, 0.111 - \frac{0.642}{\Lambda} \right
\}\,.}[]
These fits are shown in Fig.~\ref{fig:extrapolations}.
While these extrapolations should be taken with a grain of salt, it is
intriguing that the minimal point may be converging to $c(\infty) = 1/9$ or
$c(\infty)/c_{\text{free}} = 8/3$. If the minimal 4D $\CN=1$ SCFT exists
and has a simple rational central charge, this is our current best
conjecture.\footnote{If this conjecture is true, the bounds
of~\cite{Hofman:2008ar} would then imply that $\frac{1}{18} \leq a \leq
\frac{1}{6}$.} It is also possible that $\Delta_{\phi}$ is converging to the
rational value $\Delta_{\phi}(\infty) = 10/7$.

\begin{figure}[!t]
  \begin{center}
    \includegraphics{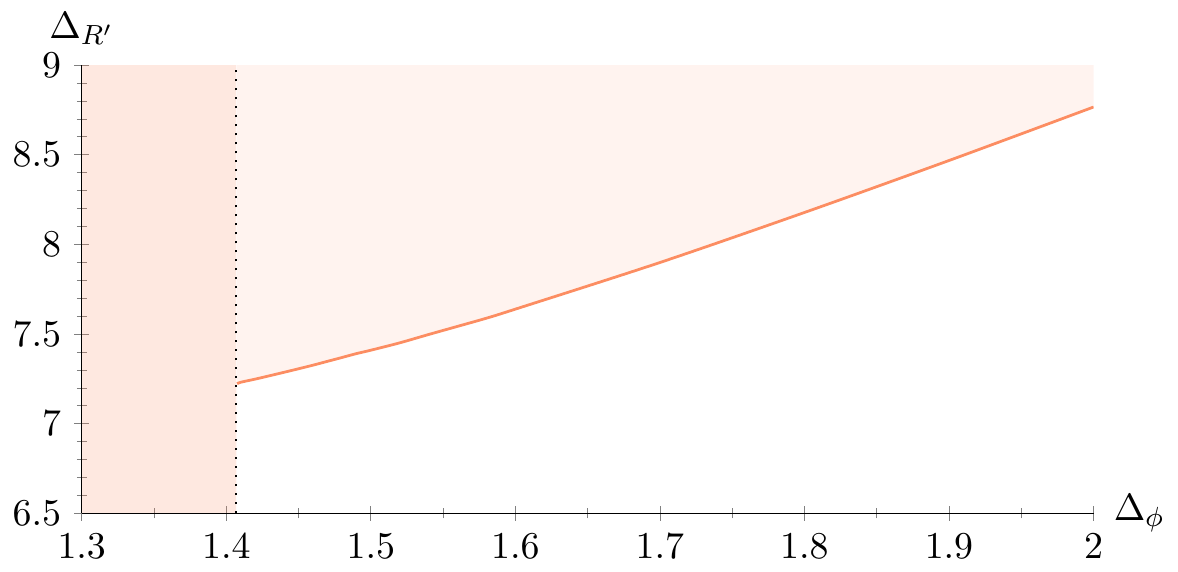}
  \end{center}
  \vspace{-12pt}
  \caption{Upper bound on the dimension of the second superconformal
  primary real scalar in the OPE $\bar{\phi}\times\phi$ as a function of
  the dimension of $\phi$, assuming that the dimension of $\bar{\phi}\phi$
  saturates its bound, i.e.\
  $\Delta_{\bar{\phi}\phi}=\Delta_{\bar{\phi}\phi}^{\text{(bound)}}$. The
  shaded area is excluded.  Everything to the left of the vertical dotted
  line at $\Delta_{\phi}=1.407$ is excluded due to the assumption that
  there is no $\phi^2$ operator.  In this plot we use $\Lambda=21$.}
  \label{fig:dim_second_real_scalar_plot_with_assumption}
\end{figure}

We finish with some preliminary explorations of the higher spectrum. In
Fig.~\ref{fig:dim_second_real_scalar_plot_with_assumption} we show the
upper bound on the dimension of the second nonchiral scalar in $\bar{\phi}
\times \phi$, assuming that the first saturates its upper bound and also
assuming the chiral ring relation $\phi^2 = 0$.
Based on this we obtain the estimate $\Delta_{R'} \lesssim 7.2$.

\begin{figure}[!t]
  \begin{center}
    \includegraphics{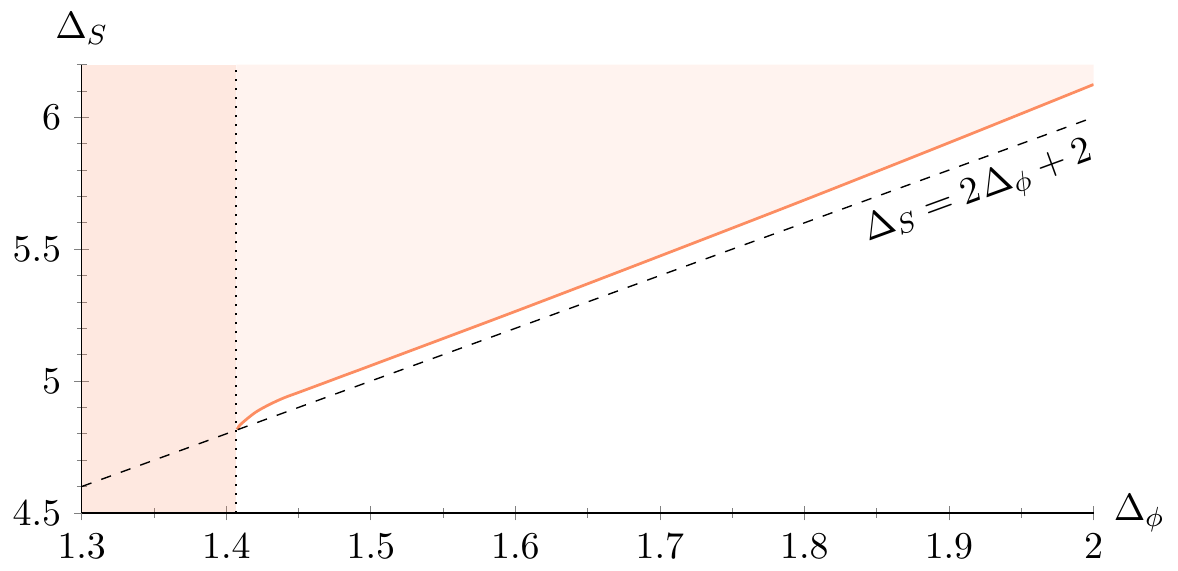}
  \end{center}
  \vspace{-12pt}
  \caption{Upper bound on the dimension of the leading superconformal
  primary spin-2 operator in the OPE $\bar{\phi}\times\phi$ as a function
  of the dimension of $\phi$. The shaded area is excluded. Everything to
  the left of the vertical dotted line at $\Delta_{\phi}=1.407$ is excluded
  due to the assumption that there is no $\phi^2$ operator. The generalized
  free theory dashed line $\Delta_S=2\hspace{0.5pt}\Delta_\phi+2$ is also
  shown.  In this plot we use $\Lambda=21$.}
  \label{fig:dim_spin_2_plot_with_assumption}
\end{figure}

In Fig.~\ref{fig:dim_spin_2_plot_with_assumption} we show an upper bound on
the leading spin-2 superconformal primary in $\bar{\phi} \times \phi$
assuming $\phi^2=0$ in the chiral ring.
At least at $\Lambda = 21$, this bound is very close to the generalized
free value when $\Delta_{\phi}$ attains its minimal value, $\Delta_S
\lesssim 4.82$. We do not know why this is the case, given that the chiral
ring relation does not hold in the generalized free solution and we could
potentially exclude this line for $\Delta_{\phi} < 3/2$.

It will be interesting in future studies to see how much of these allowed
regions are compatible with the conditions of crossing symmetry for larger
systems of correlators---in particular we would like to know whether our
minimal solution survives and can be isolated e.g.\ using the condition
that the $\phi \times \bar{\phi}\phi$ OPE contains a gap between $\phi$ and
the next scalar operator. We hope that pursuing a mixed correlator study
will lead to small islands similar to what was found
in~\cite{Kos:2014bka,Simmons-Duffin:2015qma,Kos:2015mba}.  It would also be
interesting to see if there are corresponding minimal theories with more
general chiral ring relations $\phi^n=0$. We hope to pursue these
directions in a future study.

If this solution survives, the crucial question is to identify the
underlying nature of this theory. The small central charge $c \simeq 1/9$
indicates that this theory must have a very small amount of matter and this
is not very easy to accomodate in asymptotically-free 4D gauge theories.
For example, $\CN=1$ SQCD theories all have central charge larger than 1.
The properties of this theory are similar to Wess--Zumino models with a
$W=\phi^3$ superpotential, but it has been known for a long time that such
theories do not have an interacting fixed point in 4D~\cite{Ferrara:1974fv,
Nappi:1983jw}. Thus, it may be that we have stumbled across a new
non-Lagrangian $\CN=1$ SCFT.  It would be interesting to better understand
if it could arise as a deformation of a known non-Lagrangian theory such as
one of the Argyres--Douglas fixed points~\cite{Argyres:1995jj}, or perhaps
by coupling a known $\CN=1$ SCFT to a topological field
theory~\cite{Kapustin:2014gua}. We leave further exploration of these
possibilities to future work.

\ack{We would like to thank Chris Beem, Ken Intriligator, Filip Kos,
Daliang Li, Juan Maldacena, David Simmons-Duffin, Alessandro Vichi, and Ran
Yacoby for discussions, and Sheer El-Showk, Ken Intriligator, Miguel
Paulos, David Simmons-Duffin, and Alessandro Vichi for comments on the
draft.  This research is supported by the National Science Foundation under
Grant No.~1350180. We thank the Aspen Center for Physics for hospitality
during the completion of this work, supported by NSF Grant No.~1066293. The
computations in this paper were run on the Omega and Grace computing
clusters supported by the facilities and staff of the Yale University
Faculty of Arts and Sciences High Performance Computing Center.}

\bibliography{BootstrapN1SCFTs}

\end{document}